\documentclass[aps,prb,twocolumn, preprintnumbers,superscriptaddress, showpacs]{revtex4-1}

\usepackage{graphicx}
\usepackage{dcolumn}

\begin{document}

\preprint{Ver. 6}

\title{Hysteretic superconducting resistive transition in Ba$_{0.07}$K$_{0.93}$Fe$_2$As$_2$}


\author{Taichi Terashima}
\affiliation{National Institute for Materials Science, Tsukuba, Ibaraki 305-0003, Japan}
\affiliation{JST, Transformative Research Project on Iron Pnictides (TRIP), Chiyoda, Tokyo 102-0075, Japan}
\author{Kunihiro Kihou}
\affiliation{JST, Transformative Research Project on Iron Pnictides (TRIP), Chiyoda, Tokyo 102-0075, Japan}
\affiliation{National Institute of Advanced Industrial Science and Technology (AIST), Tsukuba, Ibaraki 305-8568, Japan}
\author{Megumi Tomita}
\author{Satoshi Tsuchiya}
\author{Naoki Kikugawa}
\affiliation{National Institute for Materials Science, Tsukuba, Ibaraki 305-0003, Japan}
\author{Shigeyuki Ishida}
\author{Chul-Ho Lee}
\author{Akira Iyo}
\author{Hiroshi Eisaki}
\affiliation{JST, Transformative Research Project on Iron Pnictides (TRIP), Chiyoda, Tokyo 102-0075, Japan}
\affiliation{National Institute of Advanced Industrial Science and Technology (AIST), Tsukuba, Ibaraki 305-8568, Japan}
\author{Shinya Uji}
\affiliation{National Institute for Materials Science, Tsukuba, Ibaraki 305-0003, Japan}
\affiliation{JST, Transformative Research Project on Iron Pnictides (TRIP), Chiyoda, Tokyo 102-0075, Japan}


\date{\today}

\begin{abstract}
We have observed hysteresis in superconducting resistive transition curves of Ba$_{0.07}$K$_{0.93}$Fe$_2$As$_2$ ($T_c\sim$8 K) below about 1 K for in-plane fields.  The hysteresis is not observed as the field is tilted away from the $ab$ plane by 20$^{\circ}$ or more.  The temperature and angle dependences of the upper critical field indicate a strong paramagnetic effect for in-plane fields.  We suggest that the hysteresis can be attributed to a first-order superconducting transition due to the paramagnetic effect.  Magnetic torque data are also shown.
\end{abstract}


\maketitle



\newcommand{\ud}{\mathrm{d}}
\def\degree{\kern-.2em\r{}\kern-.3em}

\section{Introduction}

Magnetic fields destroy spin-singlet superconductivity via two different mechanisms: orbital and spin paramagnetic effects.\cite{[{For example, see }] Decroux82Book}
The former leads to the formation of a mixed state and, in the absence of the latter, a second-order transition to the normal state at the orbital critical field $B_{c2}^*(0) = \Phi_0/2\pi\xi^2$, where $\Phi_0$ and $\xi$ are the flux quantum and superconducting coherence length, respectively.
The latter lowers the normal state energy due to spin polarization and, in the absence of the former, may cause a first-order transition to the normal state at the paramagnetic critical field $B_{po} = \Delta/\sqrt{2}\mu_B$, where $\Delta$ and $\mu_B$ are the superconducting energy gap and Bohr magneton, respectively.
The Maki parameter $\alpha = \sqrt{2}B_{c2}^*(0)/B_{po}$ describes the relative importance of the two effects.
The three parameters may be estimated from experimental data using the weak-coupling BCS relations: $B_{c2}^*(0) = 0.693|B_{c2}'|T_c$, $B_{po}$ (in Tesla) =1.84$T_c$, and $\alpha=0.528|B_{c2}'|$, where $B_{c2}' = \mathrm{d}B_{c2}/\mathrm{d}T|_{T_c}$.

It was predicted that when $\alpha > 1$ the transition from the superconducting to normal state becomes first order at low temperatures.\cite{Maki64Physics1_127}
It was subsequently proposed that in such cases a novel superconducting state, now called the FFLO state, in which an order parameter oscillates in real space due to a finite center-of-mass momentum $q \neq 0$ of Cooper paris, may occur between the BCS and normal states.\cite{Fulde64PhysRev.135.A550, Larkin64zz}
Although those predictions were already made in mid 1960's, strong experimental evidence for them has appeared only recently, for two types of compounds: a heavy-fermion superconductor CeCoIn$_5$ \cite{Murphy02PRB, Bianchi02PRL, Tayama02PRB, Radovan03Nature, Bianchi03PRL} and organic superconductors.\cite{Singleto00JPCM, Tanatar02PRB, Uji06PRL, Lortz07PRL, Cho09PRB}

Iron-pnictide superconductors, first discovered by Kamihara \textit{et al}.,\cite{Kamihara08JACS} are also good candidates for the observation of the first-order transition or the FFLO state.\cite{[{For a review, see }] Gurevich11RPP}
They have large upper critical fields $B_{c2}$, and clear paramagnetic limiting of $B_{c2}(T)$ curves has indeed been reported for some of them.\cite{Hunte08Nature, Altarawneh08PRB, Yuan09Nature, Fuchs09NJP, Terashima09JPSJKFA, Kurita11JPSJ, Cho11PRB}
Importantly they are multi-band superconductors.\cite{[{For a review, see }] Johnston10AdvPhys}
In multi-band superconductors, high-field behavior due to the spin paramagnetic effect would contain much richer physics than that in single-band superconductors, as illustrated by the following questions.\cite{Gurevich11RPP}
The Maki parameter may be estimated for each band from band parameters.
If $\alpha > 1$ for only one band, does the transition become first order ?
The FFLO $q$ vector is given by $q=g\mu_BH/\hbar v_F$ in the simplest case, and hence $q$ varies from band to band.
How is a compromise reached in reality ?

We report here electrical resistance and magnetic torque measurements on Ba$_{00.7}$K$_{0.93}$Fe$_2$As$_2$.
It is a multi-band superconductor, where the Fermi surface consists of three large hole cylinders at the zone center and small hole cylinders near the corners.\cite{Sato09PRL, Terashima10JPSJ, Yoshida11JPCS, Terashima13condmatdHvA}
The size of the superconducting energy gap varies considerably from FS cylinder to cylinder.\cite{Okazaki12Science, Malaeb12PRB}
 We observe hysteresis in resistive transition curves at low temperatures for in-plane fields and suggest that it is due to a first-order superconducting  transition.

\section{Experiments}

High-quality Ba$_{0.07}$K$_{0.93}$Fe$_2$As$_2$ single crystals were prepared by a self-flux method.\cite{Kihou10JPSJ}
The Ba-to-K ratio was determined from energy-dispersive X-ray analysis (average of measurements on three pieces at 10 points per each piece).
The variation in the Ba content was within $\sim\pm$0.02 from sample to sample and also from point to point.
Standard four-contact electrical resistance $R$ measurements were performed in a dilution refrigerator and superconducting magnet.
A low-frequency ac current (typically $f$ = 13 Hz and $I$ = 100 $\mu$A, corresponding to the current density in the order of 1 A/cm$^2$) was applied in the $ab$ plane and perpendicular to the applied field $B$.
The field angle $\theta$ is measured from the $c$ axis: $\theta$ = 90$^{\circ}$ for $B \parallel ab$.
Among five measured samples from the same growth batch, the resistive hysteresis was clearly observed in two, only just observed in one, and not observed in the last two.
The clearly hysteretic samples, called H1 and H2 hereafter, have the residual resistivity ratio at $T$ = 12 K and $Tc$ of (H1) 62 and 8.2 K and of (H2) 74 and 7.4 K. 
The slight difference in $T_c$ is probably due to difference in the composition.
The two samples showing no hysteresis have similar resistivity ratios: 65 and 75, respectively.
However, the transition widths of $R(B)$ curves are very different between the two groups.
The 10\% to 90\% widths for $B \parallel ab$ at $T <$ 0.05 K are 0.95 and  0.64 T for H1 and H2, respectively, while those are 1.8 and 2.8 T for the  samples showing no hysteresis.
As shown below (Figs. 1 and 2), transition curves of the hysteretic samples, especially H2, consist of small steps.
This probably indicates that the samples are composed of a small number of domains which are homogeneous inside but have slightly different compositions from each other and that the steps correspond to respective $B_{c2}$'s of the domains. 
On the other hand, transition curves of the samples showing no hysteresis are much broader and featureless.
It seems that the intrinsic hysteresis is blurred by larger compositional inhomogeneity.
Magnetic torque measurements were performed on small pieces cut from samples H1 and H2 using piezoresistive microcantilevers.

\section{Results and discussion}

\begin{figure}
\includegraphics[width=8.6cm]{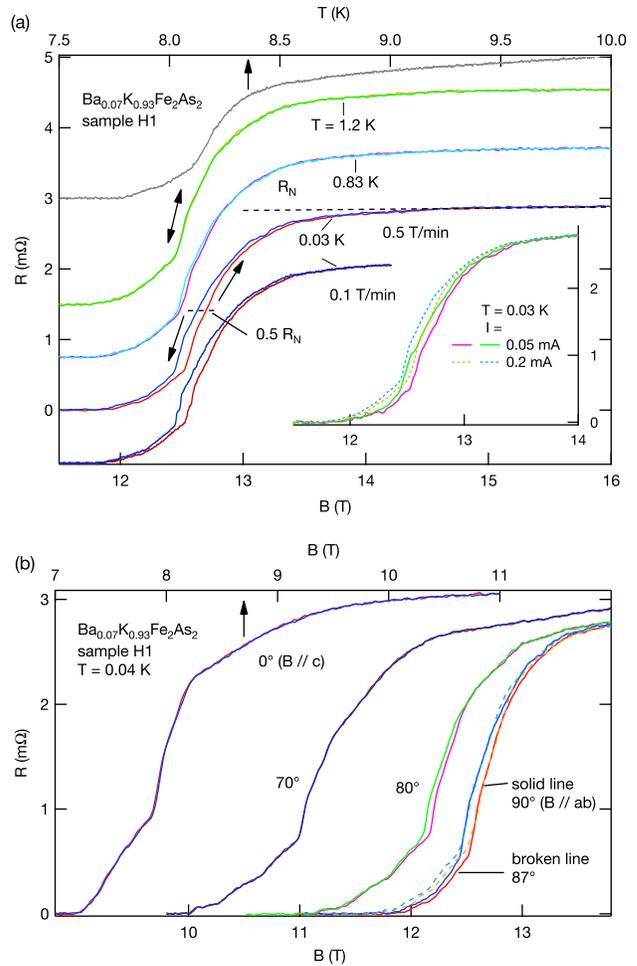}
\caption{\label{fig1}(color online).  (a) The top curve shows a zero-field superconducting transition in sample H1 as a function of temperature (top axis).  The others show resistive transitions in fields parallel to the $ab$ plane at selected temperatures as a function of field (bottom axis).  The curves are vertically shifted for clarity.  Hysteresis is visible at $T$ = 0.03 and 0.83 K but not at 1.2 K.  The two lowest curves were obtained with five times different field sweep rates.  A 50\% criterion for the determination of $B_{c2}$ is explained for the upper $T$ = 0.03 K curve.  The inset shows the current dependence of the transition curves.  (b) Resistive transition curves for selected field angles.  The hysteresis is visible at $\theta$ = 90, 87, and 80$^{\circ}$ but not at 70$^{\circ}$}   
\end{figure}

\begin{figure}
\includegraphics[width=8.6cm]{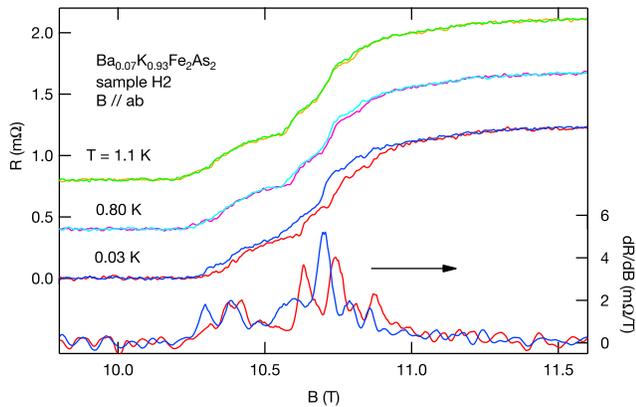}
\caption{\label{fig2}(color online).  (a) Superconducting resistive transitions in sample H2 at selected temperatures.  The curves are vertically shifted for clarity.  Hysteresis is visible at $T$ = 0.03 and 0.80 K but not at 1.1 K.  The bottom curves are the field derivative d$R$/d$B$ at 0.03 K.}   
\end{figure}

\begin{figure}
\includegraphics[width=8.6cm]{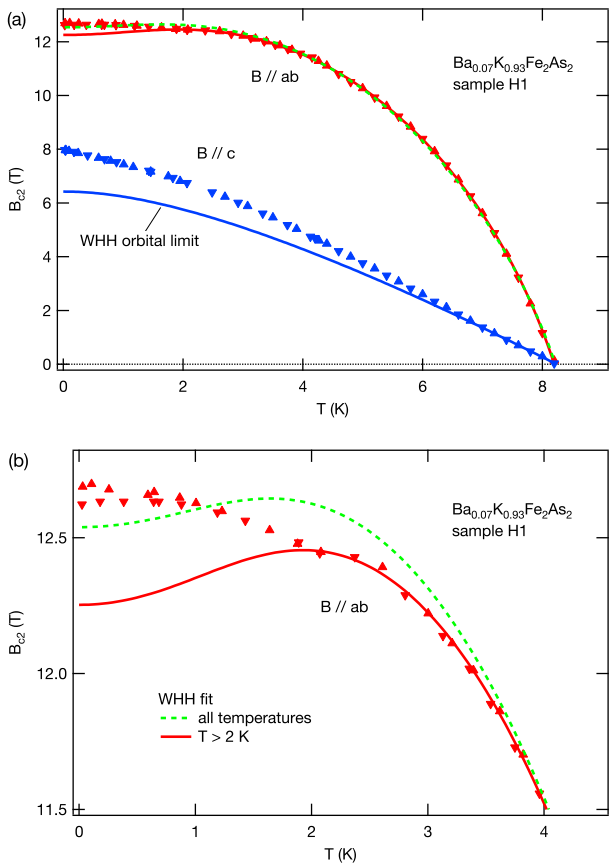}
\caption{\label{fig3}(color online)  (a) Temperature dependence of the upper critical field in sample H1 and (b) enlargement of a low-temperature part for $B \parallel ab$.  Upward and downward triangles correspond to up- and down-field data, respectively.  The dashed curve for $B \parallel ab$ is a WHH fit in the whole temperature range, while the solid one is a fit in a temperature range $T >$ 2 K.  The solid curve for $B \parallel c$ is a WHH orbital limit curve.}   
\end{figure}

\begin{figure}
\includegraphics[width=8.6cm]{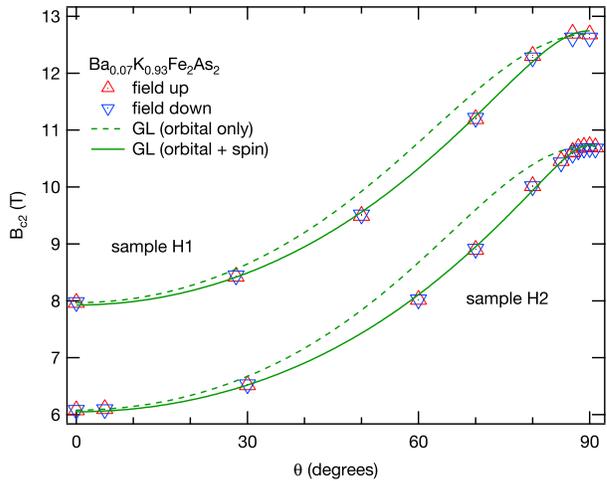}
\caption{\label{fig4}(color online)  (a) Angle dependence of the upper critical field in samples H1 and H2.  Upward and downward triangles correspond to up- and down-field data, respectively.  The dashed and solid curves are calculated with the anisotropic Ginzburg-Landau theory.  The former do not include the paramagnetic effect, while the latter do.}   
\end{figure}

Figure 1(a) shows superconducting resistive transitions in sample H1 at selected temperatures (except for the top curve).
At $T$ = 0.03 K, the hysteresis is visible between $B \sim$11.8 T, where a finite resistance appears, and $\sim$13.5 T.  
There is a kink at $\sim$12.5 T, above which the resistance rises steeply.
The zero-field transition curve (top curve) also shows a similar kink and low-temperature tail below it, which might suggest the existence of low-$T_c$ and hence low-$B_{c2}$ regions in the sample.
Five times difference in the field sweep rate (0.5 vs 0.1 T/min) leads to no essential difference in the $R(B)$ curves (compare the two lowest curves).
Four times difference in the measuring current (0.05 vs 0.2 mA) leads to no essential difference except for a slight non-Ohmic dependence (inset).
These observations seem to exclude possible extrinsic origins of the hysteresis such as history-dependent inhomogeneous current path or induced current during field ramping, which would largely be affected by the current density or sweep rate.
As the temperature is raised, the hysteresis is still visible at 0.83 K but not at 1.2 K.
Figure 1(b) shows $R(B)$ curves for different field orientations.
As the field is tilted from the $ab$ plane, the hysteresis is visible at $\theta$ = 80$^{\circ}$ but not at 70$^{\circ}$.

Resistance measurements on sample H2 give consistent results.
The resistive hysteresis is observed approximately up to $T$ = 1 K (Fig. 2) and for field orientations within $\sim 10^{\circ}$ of the $ab$ plane (data not shown).
Although the transition width is narrower than that in sample H1, transition consists of several steps as indicated by clear peaks in the derivative d$R$/d$B$ curves (Fig. 2, bottom curve).
As noted above, this can be interpreted as an indication that the sample consists of domains with slightly different compositions.

Figure 3(a) shows the temperature dependences of $B_{c2}$ in sample H1.
Linear fitting to three data points closest to $T_c$ gives the initial slopes of $B_{c2}'$ = -5.3 and -1.1 T/K for $B \parallel ab$ and $B \parallel c$, respectively, corresponding to $\alpha$ = 2.8 and 0.59.
The coherence lengths are calculated to be $\xi_{ab}$ = 7.2 nm and $\xi_c$ = 1.5 nm.
The Maki parameter $\alpha$ for $B \parallel ab$ is so large that theories developed for single-band superconductors would predict a first-order superconducting transition or the FFLO state.
Further, the clear flattening at low temperatures of the $B_{c2}(T)$ curve for $B \parallel ab$ indicates a strong paramagnetic limiting, and $B_{c2}(0)$ = 12.7 T is close to a simple estimate of the paramagnetic critical field from $T_c$, $B_{po}$ = 15.1 T.
A WWH fit \cite{Werthamer66PRB} to the $ab$-plane data in the whole temperature range gives the dashed line, which deviates upwards around $T$ = 2 K and then downwards below 1 K [Fig. 3(b)].
If we use only data points above 2 K, we obtain an excellent fit down to 2 K as shown by the solid line.
In either fit, $B_{c2}(T)$ decreases with decreasing $T$ below $T \sim$2 K.
Within the WHH theory,\cite{Werthamer66PRB} this suggests that the superconducting transition becomes first order in the low temperature region.
Experimentally, the hysteresis is observed below $\sim$1 K [Figs. 1(a) and 2].
The phase boundary below $\sim$1 K is nearly parallel to the $T$ axis, and hence it follows from the Clapeyron equation $\mathrm{d}B/\mathrm{d}T = -\triangle S/\triangle M$ that the entropy difference between the superconducting and normal phases at the phase boundary is nearly zero.
It would therefore be very difficult to see this phase boundary via heat capacity measurements

We also note that the experimental $B_{c2}(0)$ of 8.0 T for $B \parallel c$ is larger than the orbital critical field $B_{c2}^*(0)$ = 6.4 T estimated from the initial slope.
The solid line drawn for the $c$-axis data shows a $B_{c2}(T)$ curve for the WHH orbital limit without the paramagnetic effect.
The upward deviation of the experimental data indicates the importance of multi-band effects.\cite{Gurevich11RPP}

Figure 4 shows the angular dependences of $B_{c2}$ for samples H1 and H2.  Within the anisotropic Ginzburg-Landau (GL) theory, $B_{c2}^*(\theta)=B_{c2}^*(\theta=0)/\delta(\theta)$ with $\delta(\theta)=\sqrt{\cos^2\theta+\epsilon^2\sin^2\theta}$, where $\epsilon = B_{c2}^*(\theta=0)/B_{c2}^*(\theta=90^{\circ})$.\cite{Decroux82Book}
Neglecting the paramagnetic effect, namely identifying $B_{c2}^*$ with $B_{c2}$, these formulas are often used to describe the angular dependence of $B_{c2}$.
However, they fail in the present cases.
The dashed curves are drawn using the experimental values of $B_{c2}^*(\theta=0)$ and $\epsilon$.
They clearly deviate from the experimental data.
When the paramagnetic effect is important, the angular dependence of $B_{c2}$ is modified: $B_{c2}(\theta)= [B_{c2}^*(\theta=0)-aB_{c2}^2(\theta)]/\delta(\theta)$, where $a$ is a parameter describing the strength of the paramagnetic effect.\cite{Decroux82Book}
This gives excellent fits to the experimental data as shown by the solid curves, again confirming the presence of the strong paramagnetic effect.
The fitted parameters are [$B_{c2}^*(\theta=0)$, $a$, $\epsilon$] = [11.6(4), 0.058(5), 0.17(4)] and [8.5(1), 0.067(2), 0.07(2)] for samples H1 and H2, respectively, with $B_{c2}^*(\theta=0)$ and $a^{-1}$ in Tesla.
 
\begin{figure}[!]
\includegraphics[width=8.6cm]{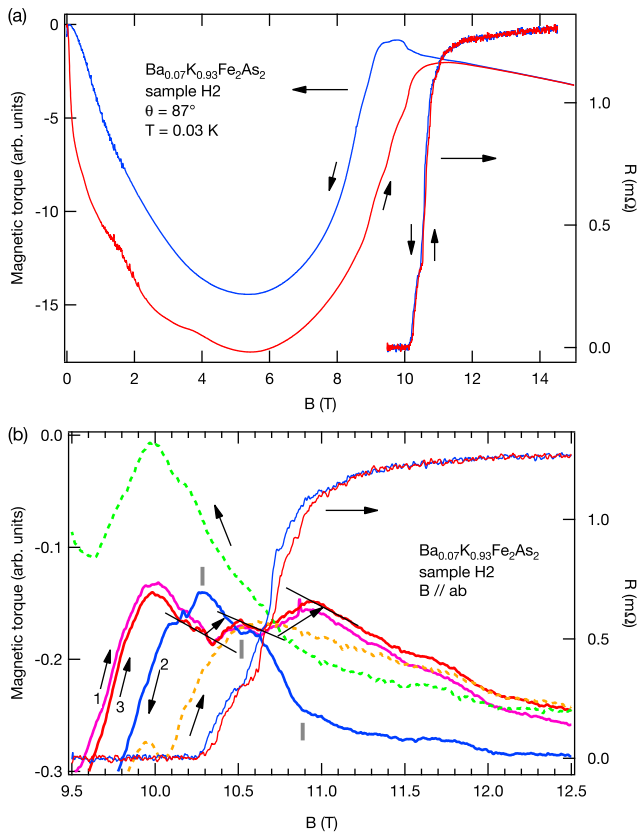}
\caption{\label{fig5}(color online)  (a) Magnetic torque in sample H2 measured at $\theta$ = 87$^{\circ}$.  The $R(B)$ curves at the same angle are shown for comparison.  (b)  Magnetic torque for $B \parallel ab$ within experimental accuracy ($\sim$0.5$^{\circ}$).  The solid and dashed curves correspond to $T$ = 0.03 and 1.5 K, respectively.  Note that the vertical scale is roughly two orders-of-magnitudes smaller than that in (a).  The $R(B)$ curves at 0.03 K for $B \parallel ab$ are shown for comparison.  Kinks appear in the torque curves at 0.03 K as indicated by the grey vertical bars attached to the field-down curve.  The kink structures may also be explicable as due to two-step jumps in the torque as illustrated with the field-up curves.  The hysteresis in the torque up to the highest shown field may be ascribed to surface superconductivity (see text).}   
\end{figure}

Finally, we show magnetic torque data measured on a small piece cut from sample H2.
Figure 5(a) shows the torque measured for the field direction $\theta$ = 87$^{\circ}$ over a wide field range.
The U-shaped torque curves with a peak approximately at $B_{c2}$/2 can be explained in the framework of the anisotropic GL theory,\cite{Hao91PRB} albeit the difference between the field-up and -down curves, which is due to vortex pinning.
The field-down curve indicates a weak peak effect in a field range $\sim$9 to 10 T.
The field-up and -down curves gradually merge above the peak effect, and the resistive transition occurs in this region as indicated by the $R(B)$ curve measured at the same angle.
The present torque curves bear a general resemblance to those reported for MgB$_2$.\cite{Fletcher04PRB}

Figure 5(b) shows the torque curves for $B \parallel ab$ within experimental accuracy ($\sim0.5^{\circ}$) measured at $T$ = 0.03 K (solid) and 1.52 K (dashed).
The $R(B)$ curve at $T$ = 0.03 K for $B \parallel ab$ is also shown for comparison.
The torque curves at $T$ = 0.03 K show kinks in the field region of the resistive transition as indicated by grey vertical bars for the field-down curve, and they approximately correspond to the onset of the resistive hysteresis, a kink in the $R(B)$ curve, and the end of the hysteresis.
The kink structures may also be interpreted as due to two successive jumps in the torque as illustrated for the field-up curves.
The kink structures are not seen in the torque curves at $T$ = 1.52 K, where the resistive transition is not hysteretic.
One might think that the observed jumps were too blunt for a first-order transition.
However, we note that magnetic torque measurements measure the component of the magnetization that is perpendicular to the field.
Magnetization jumps along the field direction could be much larger.
Indeed, it has been reported that the $ab$-plane magnetization for $B \parallel ab$ in KFe$_2$As$_2$ exhibits an abrupt increase at $B_{c2}$ at low temperatures.\cite{Burger13condmat} 

The field-up and -down curves at $T$ = 0.03 K do not merge in the shown field region, and the resistance curves are concave up to the highest shown field.
These may be ascribed to surface superconductivity, which can persist up to $B_{c3} \sim 1.7 B_{c2}$.\cite{Saint-James63PL}
Although it is difficult to precisely determine $B_{c3}$ in this sample because of drift in the measurement equipment (compare the two field-up curves 1 and 3), the torque did not show clear irreversibility beyond the drift when the field sweep direction was reversed  at 17 T from up (curve 1) to down (curve 2).
The surface superconductivity was previously reported for MgB$_2$,\cite{Lyard02PRB, Fletcher04PRB} for example.

The magnetic torque for field directions close to $B \parallel c$ exhibits de Haas-van Alphen (dHvA) oscillations for fields above $\sim$16 T, confirming the good quality of the sample.\footnote{See Supplemental Material at [URL will be inserted by publisher] for dHvA oscillations.}
Using the observed dHvA frequency of $F \approx 2000$ T, we can estimate the carrier mean free path to be $l \gtrsim50$ nm, which is much longer than $\xi_{ab}$ indicating that the sample is in the clean limit.

Theoretically, the superconducting transition in simple paramagnetically limited superconductors becomes first order below 0.56$T_c$.\cite{Maki64PTP}
Since the first order region would shrink in realistic cases with the orbital effect, the present narrow hysteretic region ($T\lesssim0.1T_c$) does not conflict with the theoretical prediction.
Because the Maki parameter $\alpha$ for $B \parallel c$ is less than one, the paramagnetic effect becomes less important as the field is tilted from the $ab$ plane, which explains the disappearance of the hysteresis for $\theta < 80^{\circ}$.
We also consider the FFLO states, since the samples are in the clean limit.
An early theoretical study predicted a possible occurrence of the FFLO state for $\alpha > 1.8$.\cite{Gruenberg66PRL}
The maximum temperature of the FFLO region was estimated to be 0.55$T_c$ for $\alpha=\infty$ but was shown to decrease with decreasing $\alpha$.
With this model,\cite{Gruenberg66PRL} the transition from the mixed state to the FFLO state was found first order, while that from the FFLO to the normal state second order.
However, recent theories predict more complicated phase diagrams, in most cases with a first-order transition line in some parts of them, and whether the FFLO-to-normal transition is first or second order depends on parameters of models considered.\cite{Burkhardt94AnnPhys, Matsuo98JPSJ, Houzet01PRB, Agterberg01JPCM, Adachi03PRB, Dalidovich04PRL, Casalbuoni04PRB, Vorontsov05PRB, Mora05PRB, Samokhin06PRB, Zhuravlev09PRB, Shimahara09PRB, Yanase09NJP}
Thus, the present observation of the resistive hysteresis does not contradict possibility of the FFLO state.

\section{Summary}

The temperature and field angle dependences of $B_{c2}$ in Ba$_{0.07}$K$_{0.93}$Fe$_2$As$_2$ indicate a strong paramagnetic effect for $B \parallel ab$, and the observed resistive hysteresis can be attributed to a first-order superconducting transition.
The features observed in the magnetic torque data may be regarded as a sign of magnetization jumps.

\end{document}